\documentclass[pra,twocolumn,showpacs]{revtex4}
\usepackage{graphicx}
\begin{document}

\bibliographystyle{unsrt}

\title{Heralding Single Photons from Pulsed Parametric Down-Conversion}
\author{T.B. Pittman, B.C Jacobs, and J.D. Franson}
\affiliation{Johns Hopkins University,
Applied Physics Laboratory, Laurel, MD 20723}

\date{\today}

\begin{abstract}
We describe an experiment in which photon pairs from a pulsed parametric down-conversion source were coupled into single-mode fibers.  Detecting one of the photons heralded the presence of the other photon in its fiber with a probability of 83\%.  The heralded photons were then used in a simple multi-photon interference experiment to illustrate their potential for quantum information applications.
\end{abstract}


\maketitle

Because the parametric down-conversion (PDC) process is known to produce pairs of photons \cite{klyshkobook,burnham70}, the detection of one member of a pair can be used to herald the presence of the other photon \cite{hong86,lvovsky01}.  For certain quantum information processing applications \cite{knill01a}, it is important to have these heralded photons in well-defined modes so that they can be used in multi-photon interference experiments with single-photons from other independent sources.  In recent years, one successful approach to this problem has been to use pulsed-pump PDC followed by narrow-band spectral filters for temporal mode definition \cite{zukowski95,bouwmeester97}, and coupling the photons into single-mode fibers for spatial mode definition.

An important experimental parameter in such a single-photon source is the heralding efficiency, $H$.  Given the detection of one member of a PDC pair, $H$ is the probability that the twin photon is actually present in its fiber.  In this paper, we describe measurements of the heralding efficiency using a  bulk-crystal Type-I PDC source pumped by pulses derived from a mode-locked laser.  These measurements are related to earlier experimental work on coupling polarization-entangled photon pairs from both continuous-wave (cw) \cite{kurtsiefer01} and pulsed \cite{bovino03} Type-II PDC bulk-crystal sources into single-mode fibers, and heralding single photons from a pulsed Type-II nonlinear waveguide PDC source in multi-mode fiber \cite{uren03}. A source of heralded PDC photons at telecom wavelengths is also being developed \cite{fasel04}. Additional studies of PDC fiber-coupling issues can be found in references \cite{monken98,castelletto03,dragan04}. 

\begin{figure}[b]
\includegraphics[angle=-90,width=3in]{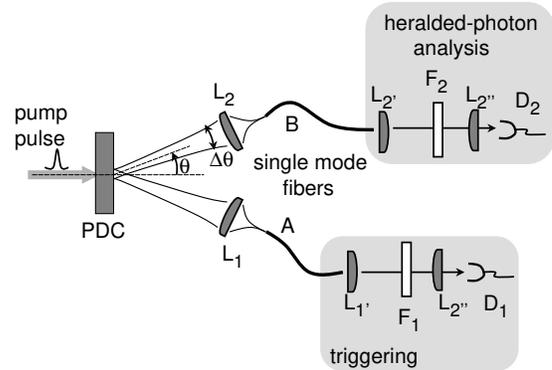}
\vspace*{-.25in}
\caption{Overview of the setup used to herald single-photons from pulsed parametric down-conversion (PDC).  The goal was to have the detection of one member of a PDC pair (the trigger photon) heralded the presence of the other photon in fiber $B$ with a high probability. Symbols are described in the text. }
\label{fig:setup}
\end{figure}

An overview of the source is shown in Figure \ref{fig:setup}.  We used a fiber-coupling strategy based on the method developed by Kurtsiefer {\em et.al.} \cite{kurtsiefer01}.  The basic idea is to image the region of the PDC crystal that is overlapped by the pump pulses into single-mode fibers, allowing for extra divergence of the beams due to the PDC phase-matching conditions over the wavelength range of interest.

In our experiment the PDC source was a 0.7 mm thick BBO crystal pumped by  frequency-doubled Ti-Sapphire laser pulses at a repetition rate of 76 MHz and an average power of 79 mW. The pump pulses were centered at 390 nm, with a duration of approximately 150 fs. The pulses were focussed to a beam diameter of roughly 0.6 mm at the location of the crystal. The BBO crystal was oriented in such a way that photon pairs at 780 nm emerged with a cone angle of $\theta = 4.5^{o}$. Two small diametrically opposed regions of the emitted cone were coupled into single-mode fibers $A$ and $B$ using aspheric lenses ($L_{1}$ and $L_{2}$) with identical focal lengths of 18.4 mm. These lenses were placed 69.4 cm from the crystal. A Gaussian beam analysis of this setup showed that photons emerging from the PDC crystal with an angular spread of roughly $\Delta\theta = 0.3^{o}$ could be coupled into the fiber cores.

The outputs of the fibers were collimated into free-space beams using additional lenses $L_{1'}$ and $L_{2'}$.  These free-space beams were then focussed onto single-photon detectors $D_{1}$ and $D_{2}$  with a final set of lenses $L_{1''}$ and $L_{2''}$.  Various spectral filters $F_{1}$ and $F_{2}$ could be inserted into these beams to study their effects on the heralding efficiency.

\begin{figure}[t]
\includegraphics[angle=-90,width=3in]{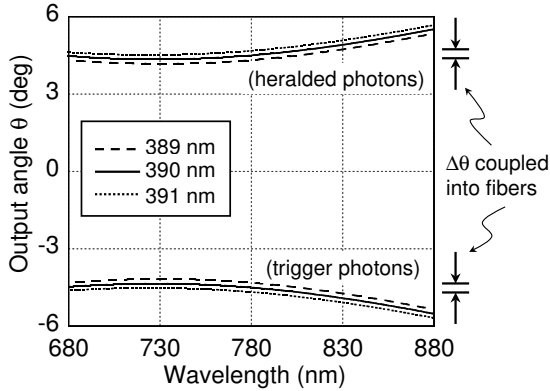}
\caption{BBO ``tuning curves'' calculated from the PDC wave-vector phase-matching condition $\vec{k}_{p}=\vec{k}_{t}+\vec{k}_{h}$, 
where $p$,$t$, and $h$ denote pump, trigger, and heralded photons. To help visualize the effects of the broad-band pumping pulses, the three sets of curves show the tuning curves for a (plane-wave) pump photon at the central wavelength of 390 nm, as well as 389 nm and 391 nm. An angular spread of roughly $\Delta\theta = 0.3^{o}$ was coupled into the single-mode fibers.}
\label{fig:tuningcurves}
\end{figure}

In contrast to the cw case, the heralding efficiency in pulsed PDC is complicated by the fact that the pumping pulses can have a bandwidth of several nanometers. One simple method for illustrating these complications is to consider the PDC ``tuning curves'' for several different values of pump wavelength.  For example, the three sets of tuning curves in Figure \ref{fig:tuningcurves} show the calculated output angle $\theta$ of the down-converted photons as a function of their wavelength for the case of a (plane-wave) pump photon at the central wavelength of 390 nm, as well as 389 nm and 391 nm. These tuning curves were calculated from the wave-vector phase matching conditions inside the crystal \cite{yarivbook}, using the Sellmeier formulae for the wavelength-dependent refractive indices of BBO.

\begin{figure}[t]
\hspace*{-.2in}
\includegraphics[angle=-90,width=4in]{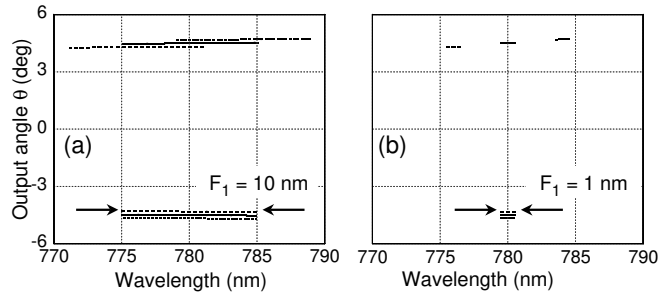}
\vspace*{-1.25in}
\caption{Zoom-in views of the tuning curves shown in Figure \protect\ref{fig:tuningcurves}. (a) shows the calculated effects of using a filter $F_{1}$ to restrict the accepted trigger photons to a bandwidth of 10 nm. Due to the broad bandwidth of the pumping pulses, the corresponding bandwidth of the heralded-photons coupled into fiber $B$ is roughly 18 nm. (b) shows the situation when filter $F_{1}$ has a narrow 1 nm bandwidth. In this case, the heralded-photon bandwidth is tailored to roughly 9 nm, which can be suitable for typical multiphoton interference experiments \protect\cite{zukowski95}.}
\label{fig:tuningcurvezoomins}
\end{figure}

From Figure \ref{fig:tuningcurves} it can be seen that the acceptance angle $\Delta\theta$ allowed down-converted photons with a raw bandwidth greater than 100 nm to be coupled into our fibers \cite{TypeIbandwidth}. Such a broad bandwidth results in heralded photons whose coherence length is too short to be used in typical multiphoton interference experiments \cite{zukowski95}. Therefore, we use a spectral filter $F_{1}$ to reduce the bandwidth of the accepted trigger photons, which causes the bandwidth of the heralded photons to also be reduced. In this way, the choice of $F_{1}$ can be used to tailor the spectral properties of the heralded photons. The effects of the broad bandwidth of the pump pulses in this scenario are illustrated in Figure \ref{fig:tuningcurvezoomins}. These plots simply zoom-in on the tuning curves of Figure \ref{fig:tuningcurves} over a smaller wavelength range.

Figure \ref{fig:tuningcurvezoomins}(a) shows the effects of restricting the accepted trigger photon bandwidth to 10nm (centered at 780 nm). Due to the broad bandwidth of the pump pulses, the corresponding heralded-photon bandwidth is nearly twice as wide.  Therefore, if an additional spectral filter $F_{2}$ is used to reduce background noise or prevent saturation of the heralded-photon detector, its bandwidth needs to be significantly wider than that of the trigger-photon filter $F_{1}$. In contrast to the cw case, the use of matched filters ($F_{1}$ and $F_{2}$) in pulsed PDC will cause an undesirable loss of heralded photons.

This effect becomes more pronounced as the accepted bandwidth of the trigger photons is further reduced. Figure \ref{fig:tuningcurvezoomins}(b) shows the case of accepting trigger photons over a wavelength range of only 1 nm.  In this case, the corresponding bandwidth of the heralded photons is tailored to nearly 10 nm. We concentrated on this case because, roughly speaking, it results in heralded-photons with coherence lengths above the threshold needed for non-classical multiphoton interference experiments using the techniques of reference \cite{zukowski95}.

\begin{figure}[b]
\includegraphics[angle=-90,width=3in]{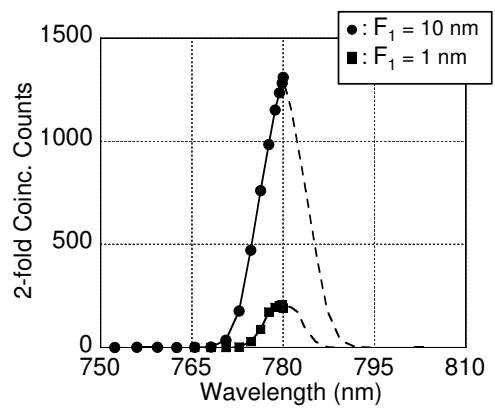}
\caption{Measured spectrum of the heralded photons when the bandwidth of the trigger photons was restricted by filter $F_{1}$ to 10 nm (solid circles) or 1 nm (solid squares). Because the simple spectrometer could only measure wavelengths lower than 780 nm, the dashed lines simply represent the mirror image of the real data points.}
\label{fig:bandwidths}
\end{figure}

The data shown in Figure \ref{fig:bandwidths} provides experimental support of the intuitive model of fiber-coupled pulsed PDC illustrated in Figure \ref{fig:tuningcurvezoomins}.  For these measurements we placed an interference filter $F_{1}$ with a bandwidth of either 10 nm or 1 nm in front of the trigger detector $D_{1}$, and passed the heralded photons through a spectrometer that consisted of an interference filter $F_{2}$ with a bandpass of 2 nm centered at 780 nm.  By twisting $F_{2}$ away from normal incidence, the filter transmission shifted to lower wavelengths, which were calibrated using a narrowband tunable diode laser.

The data of Figure \ref{fig:bandwidths} shows the average number of conditional detection events (eg. 2-fold coincidence counts between $D_{1}$ and $D_{2}$) per second, as a function of the central wavelength of the heralded photons. In comparison with the simplified model of Figure \ref{fig:tuningcurvezoomins}, the  lineshapes in Figure \ref{fig:bandwidths} are complicated by a number of factors, including the response and limited resolution of the spectrometer, the non-step shape transmission of $F_{1}$, and the pump pulse spectral profile. Nonetheless,  the broadened bandwidths of the heralded photons are evident.

The heralding efficiency $H$ was obtained by measuring the conditional detection efficiency of heralded photons, and then correcting for the losses in the heralded-photon analysis zone (the upper shaded area shown in Figure \ref{fig:setup}).  The conditional detection efficiency $\eta_{D}$ was simply defined as the coincidence counting rate, $R_{c}$, divided by the trigger photon detection rate, $R_{1}$. 

With a 1 nm bandpass filter $F_{1}$ in front of the trigger detector, and a fixed 1 nm bandpass filter $F_{2}$ in the heralded-photon analysis zone, we measured an average trigger detector rate $R_{1}$ = 3,068 Hz, and an average coincidence counting rate of $R_{c}$ = 139 Hz. This gave a conditional detection efficiency of only  $\eta_{D}$ = 4.5\%. As illustrated in Figures \ref{fig:bandwidths} and \ref{fig:tuningcurvezoomins}(b), in this ``matched-filter'' case the majority of heralded photons exiting fiber $B$ were simply rejected by the narrow 1nm bandpass filter $F_{2}$,  resulting in a low value of $\eta_{D}$.  

By replacing $F_{2}$ with a fixed 10 nm bandpass filter, the coincidence counting rate increased to $R_{c}$= 949 Hz resulting in a much larger conditional detection efficiency of $\eta_{D}$ = 31\%.  As was expected from the data shown in Figure \ref{fig:tuningcurvezoomins}(b), the use of $F_{2}$ filters with bandwidths wider than 10nm did not increase the conditional detection efficiency any further.   

Loss in the heralded-photon analysis zone of Figure \ref{fig:setup} was due to a (manufacturer-specified) detector $D_{2}$ quantum efficiency of 63\% at 780 nm, and a measured $F_{2}$ transmission of 63\% (for the 10nm bandpass filter).  In addition, we estimate 4\% reflection loss upon exiting fiber $B$ into free-space, and an additional 2\% loss due to the four anti-reflection-coated surfaces of the two lenses $L_{2'}$ and $L_{2''}$.  The total of these losses, along with the measured conditional detection efficiency of $\eta_{D}$ = 31\%, implied a heralding efficiency of $H$= 83\%.  In other words, the detection of a trigger photon heralded the presence of the twin photon in fiber $B$ with a probability of 83\% \cite{detectorefficiency}. 

\begin{figure}[t]
\includegraphics[angle=-90,width=3.25in]{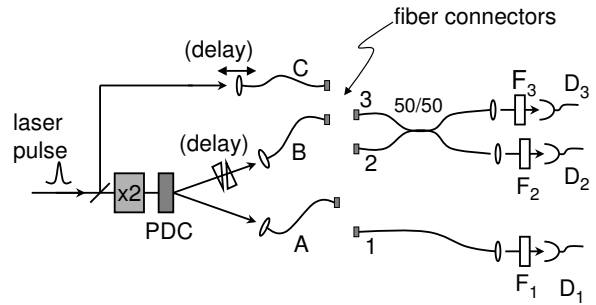}
\vspace*{-.7in}
\caption{Setup used to perform Hong-Ou-Mandel (HOM) \protect\cite{hong87} and Rarity-Tapster (RT) \protect\cite{rarity97} interference experiments with the heralded-photons in fiber $B$. For the HOM test, fibers $A$ and $B$ were connected to the inputs (labelled $2$ and $3$) of a 50/50 fiber beamsplitter.  For the RT test, the beamsplitter was fed with the heralded PDC photons from fiber $B$, and  weak coherent pulses from fiber $C$. x2 was the frequency-doubling crystal used to convert the 780 nm laser pulses into 390 nm PDC pumping pulses.}
\label{fig:beamsplitter}
\end{figure}

\begin{figure}[b]
\includegraphics[angle=-90,width=3in]{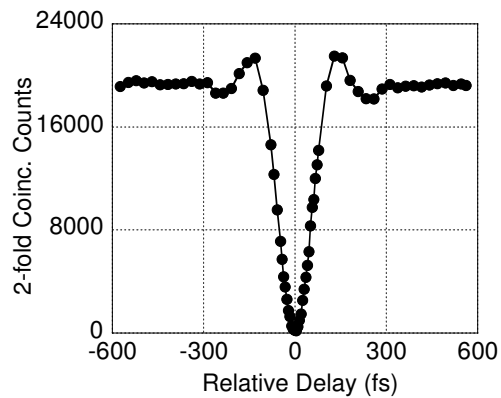}
\caption{Experimental results of a Hong-Ou-Mandel (HOM) \protect\cite{hong87} test using the PDC photon pairs. The data shows the number of coincidence counts between detectors $D_{2}$ and $D_{3}$ per 10 seconds, as a function of the relative delay between the photons. An average of 19,500 counts on the wings, compared with an average of 200 counts at the center of the dip, indicates a visibility of roughly 99\%. }
\label{fig:homdip}
\end{figure}

In order to test that all of the heralded-photons were indeed members of PDC pairs, we performed a Hong-Ou-Mandel (HOM) type interference experiment \cite{hong87}, as shown in Figure \ref{fig:beamsplitter}. Fibers $A$ and $B$ were connected to the input ports (labelled $2$ and $3$) of a 3dB (50/50) fiber coupler, and the relative delay between the photons was adjusted by inserting  translatable glass-wedges in the heralded-photon beam.  In this experiment, filters $F_{2}$ and $F_{3}$ had identical 10 nm bandwidths. Figure \ref{fig:homdip} show the resulting HOM interference dip with a visibility of roughly 99\%, which implies a negligible number of undesirable noise photons being detected.

Finally, we performed a Rarity-Tapster (RT) type experiment \cite{rarity97,pittman03a} to illustrate the use of the heralded-photons in a multiphoton interference experiment \cite{zukowski95}, as would be required in certain quantum information applications \cite{knill01a}. The RT experiment is essentially a gated-HOM beamsplitter experiment, with one of the impinging photons being the heralded PDC photon, and the second impinging photon obtained from a weak coherent state pulse. As shown in Figure \ref{fig:beamsplitter}, the weak coherent state pulses were picked off from the original 780 nm laser beam, and coupled into a third fiber $C$.

\begin{figure}[t]
\includegraphics[angle=-90,width=3in]{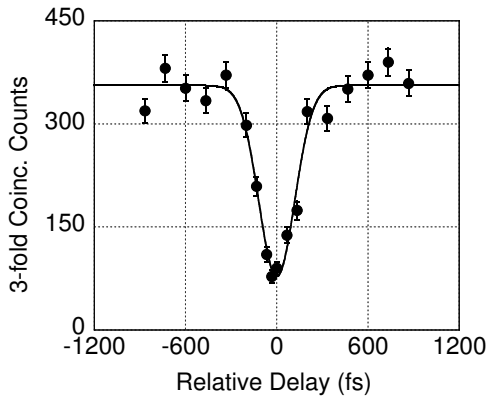}
\caption{Experimental results of a Rarity-Tapster \protect\cite{rarity97} beamsplitter test using the heralded PDC photons, and photons from a weak coherent state. The data shows the number of three-fold coincident counts between detectors $D_{1}$, $D_{2}$, and $D_{3}$ per 600 seconds.  The RT interference dip was fit by a simple Gaussian function with a visibility of 78\%. }
\label{fig:rtdip}
\end{figure}

Figure \ref{fig:rtdip} shows the resulting RT interference dip as a function of the delay of the photons in fiber $C$ relative to the heralded PDC photons in fiber $B$.  For these measurements, the trigger filter $F_{1}$ again had a 1 nm bandwidth, while $F_{2}$ and $F_{3}$ had 10 nm bandwidths.  The visibility of 78\% in this simple test can be viewed as a measure of the indistinguishability of the photon sources.

The motivation for this work was within the context of a reliable periodic source of single-photons based on stored PDC \cite{pittman02a}. The basic idea there is to use the detection of a trigger photon to actively switch the heralded-photon into a fiber storage loop.  The stored photon is then known to be circulating in the loop, and can then be switched out when needed.  The two main technical challenges in that type of periodic source are minimizing losses in the switch, and maximizing the heralding efficiency $H$.  These issues are also crucial in other single-photon sources based on PDC \cite{migdall02,jeffery04}.

In this regard, the heralding efficiency of $H$=83\% measured here using fiber-coupled photons from a fairly typical pulsed PDC source is encouraging, and their appear to be possibilities for further improvements.  In addition, a promising alternative for high heralding efficiency would be to use photon pairs created inside optical fibers \cite{fiorentino02}.  Furthermore, the use of nonlinear waveguides (rather than bulk crystals) as a PDC source is expected to greatly enhance the mode matching into single-mode fibers \cite{uren03}.

We would like to acknowledge useful discussions with Paul Kwiat, Alan Migdall, Harald Weinfurter, and Hugo Zbinden.  This work was supported by ARO, NSA, ARDA, ONR, and IR\&D funding.




\begin{thebibliography}{50}

\bibitem{klyshkobook} {\em Photons and Nonlinear Optics}, D.N. Klyshko, Gordon and Breach Science Publishers (1988).

\bibitem{burnham70} D.C. Burnham and D.L. Weinberg, Phys. Rev. Lett. {\bf 25}, 84 (1970).

\bibitem{hong86} C.K. Hong and L. Mandel, Phys. Rev. Lett. {\bf 56}, 58 (1986).

\bibitem{lvovsky01} A.I. Lvovsky {\em et.al.}, Phys. Rev. Lett. {\bf 87}, 050402 (2001).

\bibitem{knill01a} E. Knill, R. Laflamme, and G.J. Milburn, Nature {\bf 409}, 46 (2001).

\bibitem{zukowski95} M. Zukowski, A. Zeilinger, and H. Weinfurter, Ann. N.Y. Acad. Sci. {\bf 755} 91, (1995). J.G. Rarity, {\em ibid} {\bf 755}, 624 (1995).

\bibitem{bouwmeester97} D. Bouwmeester {\em et.al.}, Nature {\bf 390}, 575 (1997);


\bibitem{kurtsiefer01} C. Kurtsiefer, M. Oberparlieter, and H. Weinfurter, Phys. Rev. A {\bf 64}, 023802 (2001).

\bibitem{bovino03} F.A. Bovino {\em et.al.}, Opt. Comm. {\bf 227}, 343 (2003).

\bibitem{uren03} A.B. Uren, C. Silberhorn, K. Banaszek, and I.A. Walmsley, quant-ph/0312118 (2003).

\bibitem{fasel04} S. Fasel {\em et.al.}, quant-ph/0408136.

\bibitem{monken98} C.H. Monken, P.H. Souto Ribeiro, and S. Padua, Phys. Rev. A {\bf 57}, R2267 (1998).

\bibitem{castelletto03} S. Castelletto, I.P. Degiovanni, M. Ware, and A. Migdall,  ``{\em Quantum Communications and Quantum Imaging}'', SPIE Proceedings, R.E. Meyers and Y.H. Shih, eds., {\bf 5161}, 48 (2003).

\bibitem{dragan04} A. Dragan, quant-ph/0407113.

\bibitem{yarivbook} {\em Quantum Electronics}, A. Yariv, John Wiley Publishers, New York (1989).

\bibitem{TypeIbandwidth} The fiber-coupled bandwidth in this Type-I PDC experiment is significantly larger than that of the Type-II PDC setup in reference \protect\cite{kurtsiefer01}.

\bibitem{detectorefficiency} Relying on the manufacturer-specified value of 63\% for the quantum efficiency of $D_{2}$ was the largest uncertainty in the measurement of $H$. As an interesting side-note, if we {\em assume} perfect 
fiber-coupling of the heralded photons (aside from a 4\% loss into the uncoated fiber), the measured counting rates would imply a $D_{2}$ quantum efficiency of 55\% using an absolute detector-callibration method (for details, see A.  Migdall, Physics Today {\bf 52}, 41 (1999)).   

\bibitem{hong87} C.K. Hong, Z.Y. Ou, and L. Mandel, Phys. Rev. Lett. {\bf 59}, 2044 (1987).

\bibitem{rarity97} J.G. Rarity and P.R. Tapster, Philos. Trans. R. Soc. London {\bf A 355}, 2267 (1997); quant-ph/9702032.

\bibitem{pittman03a} T.B. Pittman and J.D. Franson, Phys. Rev. Lett. {\bf 90}, 240401 (2003).   

\bibitem{pittman02a} T.B. Pittman, B.C. Jacobs, and J.D. Franson, Phys. Rev. A {\bf 66}, 042303 (2002).

\bibitem{migdall02}  A.L. Migdall, D.Branning, and S.Castelletto,  
Phys. Rev. A {\bf 66}, 053805 (2002).

\bibitem{jeffery04}  E. Jeffrey, N. Peters, and P. G. Kwiat,  New J. Phys. {\bf 6}, 100 (2004).

\bibitem{fiorentino02}
M. Fiorentino, P.L. Voss, J.E. Sharping, and P.Kumar, {\em IEEE Phot. Tech. 
Lett.} {\bf 14}, 983 (2002).







\end{thebibliography}
\end{document}